\documentclass[aps,prb,twocolumn,showpacs,superscriptaddress]{revtex4-1}

\usepackage{graphicx}
\usepackage{dcolumn}
\usepackage{bm}
\usepackage{amssymb}
\usepackage{colortbl}
\usepackage{dsfont}

\begin{document}

\title{
Entanglement discrimination in multi-rail electron-hole currents
}

\author{J. P. Baltan\'as}
\email{baltanas@us.es}
\affiliation{Departamento de F\'{\i}sica Aplicada II, Universidad de Sevilla, E-41012 Sevilla, Spain}
\author{D. Frustaglia}
\email{frustaglia@us.es}
\affiliation{Departamento de F\'{\i}sica Aplicada II, Universidad de Sevilla, E-41012 Sevilla, Spain}

\date{\today}


\begin{abstract}
We propose a quantum-Hall interferometer that integrates an electron-hole entangler with an analyzer working as an entanglement witness by implementing a multi-rail encoding. The witness has the ability to discriminate (and quantify) spatial-mode and occupancy entanglement. This represents a feasible alternative to limited approaches based on the violation of Bell-like inequalities.
\end{abstract}



\maketitle


\section{Introduction} 

The reliable production and detection of quantum-entangled electron currents is a relevant issue in the roadmap towards solid-state quantum information based upon flying qubits. To this aim, several schemes have been settled along the last years. For the production, the most noticeable proposals rely on Cooper pair emission from superconducting contacts, \cite{RSL01,LMB01} correlated electron-hole pair production in tunnel barriers, \cite{BEKV03} and integrated single-particle emitters. \cite{SMB09} It is a widespread belief that these mechanisms are likely to produce highly entangled electron currents. Unfortunately, serious difficulties arise for the detection and quantification of the entanglement produced by those means. Ideal approaches are based on the violation of Bell-like inequalities \cite{BCHSH} in terms of zero-frequency noise correlators, \cite{SSB03} sometimes including postselection mechanisms. However, corresponding efforts have been unsuccessful so far, most probably due to technological limitations for the controlled manipulation of a relatively large amount of parameters (only indirect signatures of entanglement as the two-particle Aharonov-Bohm effect \cite{SSB04} have been found in the laboratories \cite{NOCHMU07}). In this situation, alternative approaches were developed with a more pragmatic viewpoint in the form of entanglement witnesses, namely, specific observables that can detect entangled states belonging to a certain subspace of interest by introducing a limited amount of controlled parameters. \cite{witness} In particular, a series of works has addressed the possibility of bounding the entanglement of electronic currents via single observables by mapping the probe states into Werner states \cite{BL03, GFTF-06,GFTF-07} (a concept recently extended to isotropic states \cite{STFG11}). Moreover, Giovannetti \emph{et al.} \cite{GFTF-07} introduced a procedure allowing the discrimination of different types of entanglement, viz. the conventional mode entanglement (focused on the internal degrees of freedom of given particles) and the less appraised occupancy entanglement (relying on fluctuating local particle numbers).\cite{Wis03-Schuch04}

\begin{figure}
\includegraphics[width=\columnwidth]{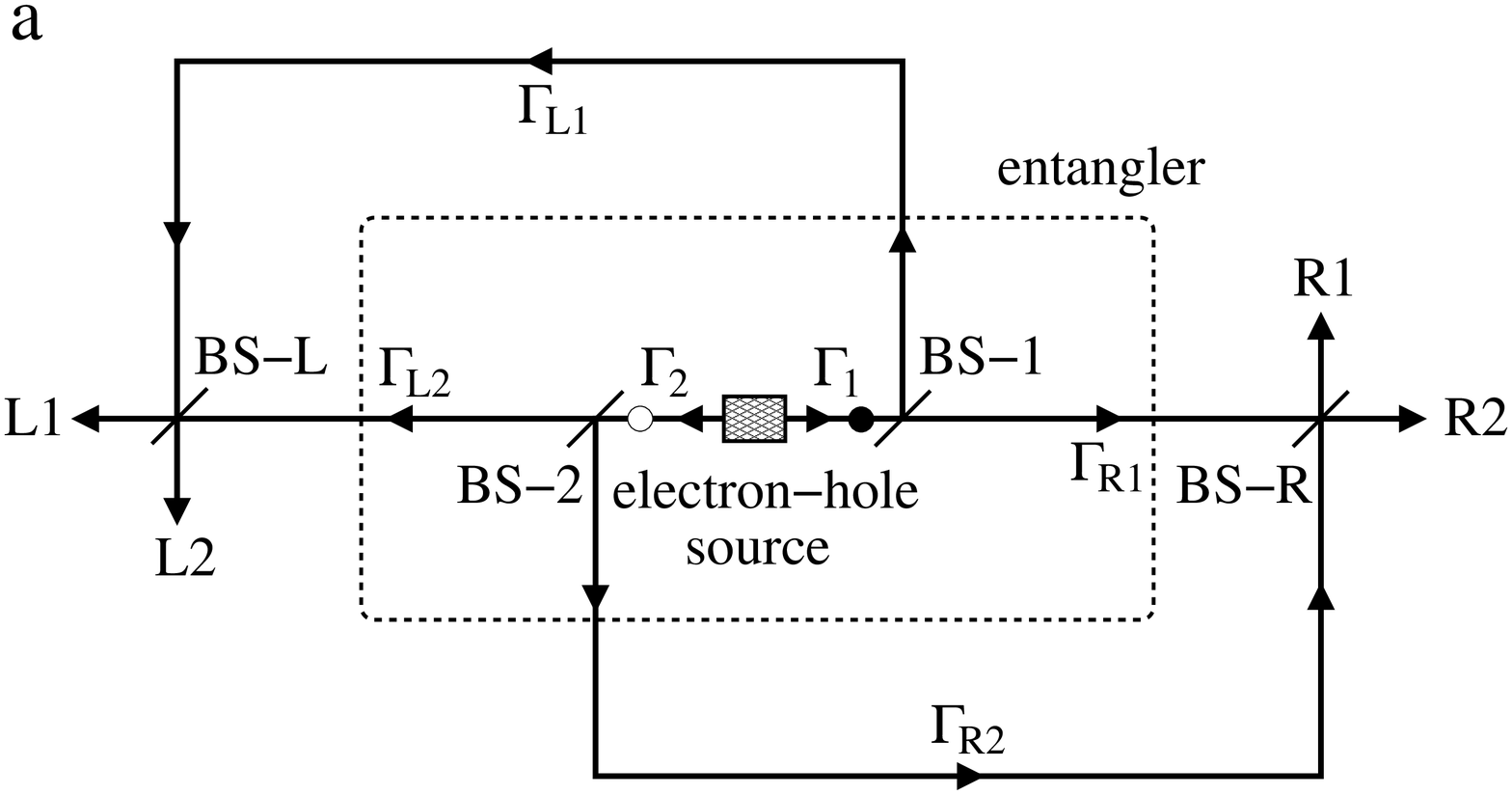}
\includegraphics[width=\columnwidth]{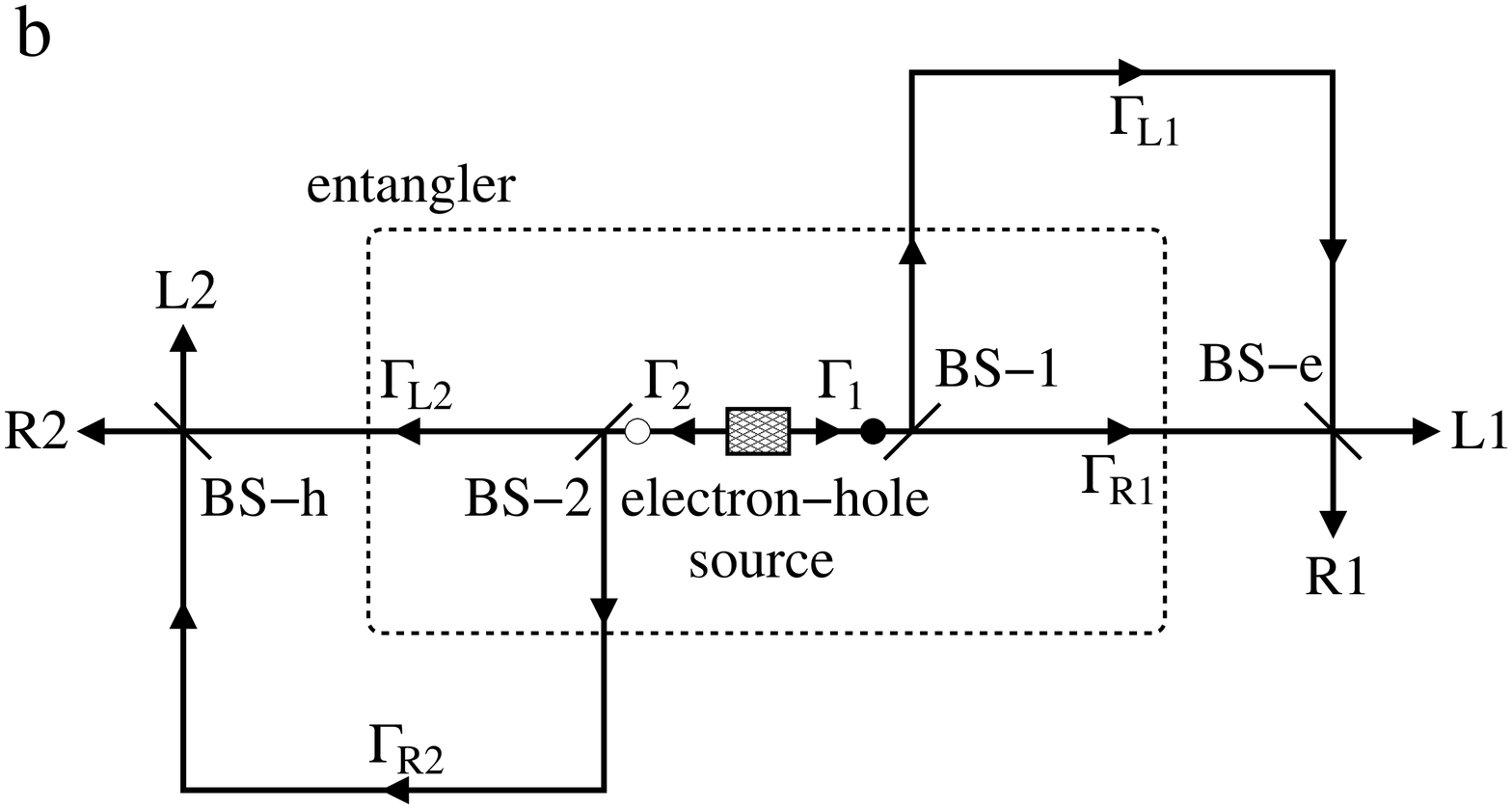}
\caption{Interferometer schemes integrating a two-particle (viz. electron-hole) multi-rail entangler (dashed square) with two alternative entanglement analyzers: a Bell-CHSH analyzer (panel a, proposed in Ref.~\onlinecite{FC-09}) for postselected spatial-mode entanglement, and a witness analyzer (panel b, proposed here based on Ref.~\onlinecite{GFTF-07}) able to discriminate spatial-mode and occupancy entanglement. The entanglement between L and R rails produced after scattering at BS-1 and BS-2 is exactly the same in both cases (dashed box). Notice that the geometry of panel b corresponds to the optical interferometer originally proposed by Franson. \cite{F-89} However, the correlations studied here differ fundamentally from Franson's (see text for details).  
}
\label{fig-1}
\end{figure}

Here, we propose an integer quantum-Hall interferometer that integrates in a single circuit an electronic entangler with an entanglement-witness analyzer. The entangler is based on a previous scheme by Frustaglia and Cabello\cite{FC-09} that employs the multi-rail encoding (i.e., spatially separated transmission channels or modes) of electron-hole pairs produced at a tunnel barrier, originally designed for exploiting only spatial-mode qubit entanglement (also referred to as path- or orbital-mode entanglement) by postselection, i.e., by filtering out any contribution to occupancy entanglement. The interferometer introduced in Ref.~\onlinecite{FC-09}, sketched in Fig.~\ref{fig-1}a, is the electronic version of the optical interferometer proposed by Cabello \emph{et al.}\cite{CRVMM09} to solve a fundamental deficiency present in the original Franson's Bell-like proposal with energy-time entanglement \cite{F-89} due to the actual existence of a local hidden variable model reproducing the observed results.\cite{AKLZ99} Our detection strategy, instead, implements the results obtained in Ref. \onlinecite{GFTF-07} for the discrimination of spatial-mode and occupancy entanglement. This relies on the inspection of current cross correlations at the output ports of an electronic beam splitter (BS), provided that controllable, channel-dependent phase shifts are introduced in one of the input ports. Notice the contrast with the approach originally adopted in Ref. \onlinecite{FC-09} based on the violation of Bell-Clauser-Horne-Shimony-Holt (CHSH) inequalities \cite{BCHSH} by noise cross correlations, \cite{SSB03,SSB04,Beenakker-06} which results inappropriate for the analysis of occupancy entanglement due to superselection rules (SSRs) induced by local particle-number conservation.\cite{Wis03-Schuch04,Beenakker-06}

\section{Electronic interferometer}

Let us start by reviewing the electronic interferometer introduced in
Ref.~\onlinecite{FC-09}, schematically depicted in Fig.~\ref{fig-1}a. A
source emits an electron-hole pair, with the electron and hole
travelling in opposite directions towards beam splitters BS-1 and
BS-2, respectively. After meeting the corresponding beam splitter,
each member of the electron-hole pair splits independently into two
paths $\Gamma_{{\mathrm R1}}/\Gamma_{\mathrm L1}$ (electron) and
$\Gamma_{\mathrm R2}/\Gamma_{\mathrm L2}$ (hole). Path
$\Gamma_{\mathrm R1}$ ($\Gamma_{\mathrm R2}$) takes the electron (hole) to
the right (R) side of the interferometer for detection, while path
$\Gamma_{\mathrm L1}$ ($\Gamma_{\mathrm L2}$) does likewise in the
left (L) side.

After scattering at BS-1 and BS-2, the resulting electron-hole
excitation consists of a multi-rail superposition containing different
number of excitations in the L and R arms of the interferometer. More
precisely, two contributions are found in which one excitation flies
off to the right and the other one to the left, together with two
contributions in which both particles fly off to the same side of the
interferometer. \cite{FC-09} When written in the \emph{left-right}
(L-R) \emph{bipartition basis}, this state displays
hyper-entanglement, viz., standard spatial-mode entanglement and
occupancy entanglement. The first one corresponds to the two-qubit
entanglement between L and R propagating channels, with exactly one
particle occupying one of the L and R channels, while the second one
results form the coherent superposition of terms with different local
particle number: two particles occupying both of the L propagating
channels (and no particles on the R channels) or vice versa
\cite{Wis03-Schuch04} (see also Ref. \onlinecite{GFTF-07}).

\begin{figure}
\includegraphics[width=0.85\columnwidth]{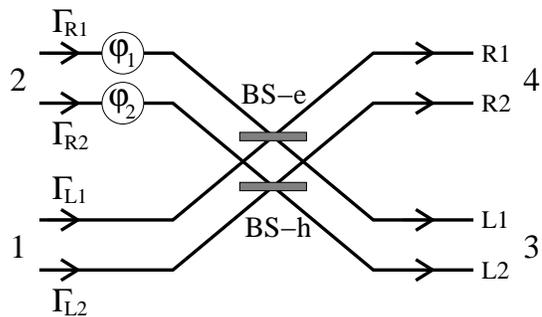}
\caption{ 
Sketch of the entanglement-witness analyzer. The beam splitter BS-e operates on rails occupied by purely electronic excitations while BS-h does similarly on hole ones. This is equivalent to a single beam splitter processing two input channels per port which remain unmixed in the scattering process. \cite{GFTF-07} Particles entering port 2 (R channels) experience an additional, channel-dependent phase ($\varphi_1,\varphi_2$) before scattering at the corresponding beam splitter. These phases are controllable via voltage gates (e.g., top or side gates) inserted along paths $\Gamma_{\mathrm R1}$ and $\Gamma_{\mathrm R2}$. The entanglement witness is built from current cross correlations at output ports 3 (terminals L1 and L2) and 4 (terminals R1 and R2). See  Fig.~\ref{fig-3} for its quantum-Hall implementation.
}
\label{fig-2}
\end{figure}

In Ref.~\onlinecite{FC-09}, postselected spatial-mode qubit entanglement is detected via the
violation of a Bell-CHSH inequality, with BS-L and BS-R playing the
role of controllable local operators acting on L and R
propagating channels. To this end, coincidence measurements used in
the optical version of the interferometer \cite{CRVMM09} are
replaced by zero-frequency current-noise cross
correlations between terminals placed at different sides of the electronic interferometer: one on the left 
(terminals L1/L2) and the other one on the right (terminals
R1/R2). By construction, this procedure postselects the components contributing to spatial-mode entanglement \emph{only}, disregarding those contributions carrying more than one excitation to the
L or R terminals. \cite{FC-09} Moreover, the use of Bell-CHSH inequalities to detect occupancy
entanglement would require local mixing of quantum states with different number of fermions, which is forbidden by SSRs.

For this reason, we reconfigure the interferometer depicted in Fig.~\ref{fig-1}a to implement the witness scheme introduced in Ref.~\onlinecite{GFTF-07} in order to address \emph{both} the spatial-mode and occupancy entaglement of two-particle probe states. This is based on the study of current cross correlations at the output ports of a BS as a function of controllable phase shifts. To this aim, we identify the L and R propagating paths in Fig.~\ref{fig-1}a, $\Gamma_{\mathrm  L1}$/$\Gamma_{\mathrm L2}$ and $\Gamma_{\mathrm  R1}$/$\Gamma_{\mathrm R2}$, with input ports 1 and 2 in the entanglement analyzer of Ref.~\onlinecite{GFTF-07} (see Fig.~\ref{fig-2}). It is most important to realize that the analyzer introduced in Ref.~\onlinecite{GFTF-07} consists in a \emph{single} BS which does not produce channel mixing in the scattering process. In our case, this is equivalent to introduce two spatially separated BSs: one for electrons (BS-e) and one for holes (BS-h), as depicted in Fig.~\ref{fig-2}. This results in the design sketched in
Fig.~\ref{fig-1}b. It is worth noting that the topology of this setup coincides with that of Franson's interfereometer. \cite{F-89} However, the correlations considered by Franson are essentially different from those needed here. Franson's setup, as demanded by Bell-CHSH tests, would require the study of noise cross correlations between L1/R1 and L2/R2 terminals in Fig.~\ref{fig-1}b (namely, between electron and hole excitations). Here, instead, in order to discriminate mode from occupancy entanglement by following the protocol described in Ref. \onlinecite{GFTF-07}, one needs the current cross correlations between output ports 3 and 4 in the analyzer of Fig.~\ref{fig-2} corresponding to terminals L1/L2 and R1/R2 (see Fig.~\ref{fig-1}b), respectively. This procedure implements nonlocal operations in the original L-R bipartition necessary for the entanglement discrimination.

To complete our device, we need to insert additional voltage gates (e.g., top or side gates) along paths
$\Gamma_{\mathrm R1}$ and $\Gamma_{\mathrm R2}$ allowing the introduction of 
\emph{controllable} phases $\varphi_{1}$ and $\varphi_{2}$ as demanded by the analyzer of Fig.~\ref{fig-2}. The resulting quantum-Hall setup is shown in Fig.~\ref{fig-3}. Electrons propagate from sources $1$ and $2$
(subject to equal voltages $V$) to grounded drains L1, L2, R1 and R2, along single-mode edge channels. Electrically controled quantum point contacts labelled as BS-$n$, with
$n=0,1,2,\mathrm{e},\mathrm{h}$, act as beam splitters. As discussed
in Ref.~\onlinecite{FC-09}, the production of entanglement involves only beam splitters BS-$0$, BS-$1$ and BS-$2$ together with the \emph{primary} source 1. The BS-0 is set to be
low transmitting (tunneling regime). Thus, an electron traveling from
primary source 1 can tunnel through BS-0 to the right side of the
interferometer, leaving a hole in the Fermi sea traveling towards the left side. After
emission at BS-0, each member of the electron-hole pair splits into a
pair of paths at BS-1 and BS-2, respectively, which results in a spatial-mode and occupancy entangled electron-hole excitation in the original L-R bipartition, as discussed above.\cite{note-1} The \emph{secondary} source 2 is not directly involved in the production of entanglement itself. Its role is to avoid
contamination of the signal generated at BS-0 with the undesired
current-noise correlations that would originate at BS-2 in the absence
of this secondary source. Entanglement is discriminated
by current cross-correlations between the output terminals of BS-e and BS-h, as mentioned above. Details are presented in the following sections.

\section{Entanglement production}

\begin{figure}
\includegraphics[width=\columnwidth]{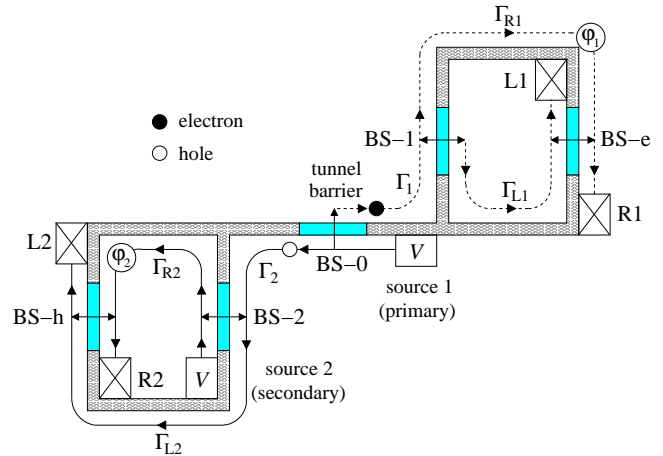}
\caption{ 
Proposed quantum-Hall setup. Solid lines represent electron streams while dashed lines correspond to empty electron channels. Electrons circulate from primary and secondary sources towards drains L1, L2, R1 and R2. Quantum point contacts BS-$n$ ($n=0,1,2,$e,h) work as electron beam splitters. The low-transparency BS-0 is a tunnel barrier producing correlated electron-hole pairs propagating in different directions that become hyper-entangled in the L-R  rail bipartition after scattering at BS-1 and BS-2. The BS-e and BS-h are the building blocks of the entanglement analyzer together with the phases $\varphi_1$ and $\varphi_2$ controlled by voltage gates. Entanglement is discriminated by studying the current cross correlations defined in Eq. (\ref{C34}).
}
\label{fig-3}
\end{figure}

We start by writing an explicit expression of the state produced at the entangler, to
be probed by the analyzer. To this end, we follow the steps detailed in
Ref.~\onlinecite{FC-09} adapted to the setup of Fig.~\ref{fig-3}. Electrons are injected from sources $1$ and $2$ with energy $\varepsilon$ on an energy window $eV$ above the Fermi sea
$|0\rangle$. Upon tunneling of electrons from source 1 (transmission
probability $T_{0}=|t_{0}|^2\ll 1$), an electron-hole pair packet is
generated at BS-0. After scattering at BS-1 and BS-2 (with
amplitudes $t_{1}$, $r_{1}$ and $t_{2}$, $r_{2}$, respectively), the pair state  
evolves into
\begin{equation}
\label{input}
|\Psi\rangle=|\bar{0}\rangle+|\bar{\Psi}\rangle,
\end{equation}
where
\begin{eqnarray}
\label{Psi_bar}
&|\bar{\Psi}\rangle&=t_{0}e^{i(\phi_{1}-\phi_{2})}\int_{0}^{eV}{\mathrm d}\varepsilon \big[t_{1}t_{2}^{\ast}e^{i(\gamma_{1}-\delta_{2})}a^{\dagger}_{\mathrm L1}(\varepsilon)a_{\mathrm R2}(\varepsilon)\nonumber\\
&-&r_{1}r_{2}^{\ast}e^{i(\delta_{1}-\gamma_{2})}a_{\mathrm L2}(\varepsilon)a^{\dagger}_{\mathrm R1}(\varepsilon)+t_{1}r_{2}^{\ast}e^{i(\gamma_{1}-\gamma_{2})}a^{\dagger}_{\mathrm L1}(\varepsilon)a_{\mathrm L2}(\varepsilon)\nonumber\\
&+&r_{1}t_{2}^{\ast}e^{i(\delta_{1}-\delta_{2})}a^{\dagger}_{\mathrm R1}(\varepsilon)a_{\mathrm R2}(\varepsilon)\big]|\bar{0}\rangle
\end{eqnarray}
describes an electron-hole wavepacket out of a redefined vacuum
$|\bar{0}\rangle=\prod_{\varepsilon=0}^{eV}a^{\dagger}_{\mathrm
  L2}(\varepsilon)a^{\dagger}_{\mathrm
  R2}(\varepsilon)|0\rangle$. Here, $a^{\dagger}_{n}$ ($a_{n}$)
creates an electron (hole) propagating towards terminal $n={\mathrm
  L1},{\mathrm L2},{\mathrm R1},{\mathrm R2}$ when BS-e and BS-h are
closed. In addition, $\phi_{m}$, $\delta_{m}$ and $\gamma_{m}$
($m=1,2$) are the phases adquired by an electron travelling along
paths $\Gamma_{m}$, $\Gamma_{{\mathrm R}m}$ and $\Gamma_{{\mathrm
    L}m}$, respectively, when the controllable phases $\varphi_{1}$
and $\varphi_{2}$ are set to $0$. The redefined vacuum
$|\bar{0}\rangle$ consists of a noiseless stream of electrons emitted
from BS-2 towards terminals L2 and R2. This is possible by virtue of 
secondary source 2. Otherwise, electrons entering BS-2 from primary source 1 alone would scatter as correlated noisy currents that mask the signatures of the
electron-hole excitations emitted from BS-0. \cite{FC-09}

Notice that the first two terms within brackets in Eq.~(\ref{Psi_bar})
correspond to a coherent superposition of an electron and a hole
entering the analyzer of Fig.~\ref{fig-2} from different ports
(two spatial-mode entangled qubits), while the last two terms describe an
electron and a hole occupying the same port when entering the analyzer (occupancy entangled excitations). In the next section, we examine both types of entanglement by following the lines established in Ref.~\onlinecite{GFTF-07} via current cross correlators.

\section{Entanglement detection}

According to Ref.~\onlinecite{GFTF-07}, both BS-e and BS-h are set to be $50\%$ beam splitters with scattering matrices given by
\begin{equation}
\hat{\bf s}^{(\alpha)}=\sqrt{1/2}\left(\begin{array}{cc}1 & e^{i\varphi_{\alpha}} \\ 1 & -e^{i\varphi_{\alpha}}\end{array}\right),
\end{equation}
with $\alpha=1,2$ for BS-e and BS-h, respectively. These matrices relate the annihilation operators on both sides of the analyzer of Fig.~\ref{fig-2} as  
${\bf b}_{(\alpha)}(\epsilon)=\hat{\bf s}^{(\alpha)} {\bf a}_{(\alpha)}(\epsilon)$, with incoming ${\bf a}^\dag_{(\alpha)}=(a\dag_{R\alpha},a^\dag_{L\alpha})$ and outgoing ${\bf b}^\dag_{(\alpha)}=(b^\dag_{R\alpha},b^\dag_{L\alpha})$.

The keystone for entanglement detection are the correlations of electron and hole excitations at output ports $3$ and $4$ in Fig.~\ref{fig-2}. These are properly accounted by the dimensionless current cross correlator\cite{GFTF-07}
\begin{equation}
\mathcal{C}_{34}=\frac{h^2\nu^{2}}{2e^{2}}\lim_{T\to\infty}\int_{0}^{T}dt_{1}dt_{2}\frac{\langle I_{3}(t_{1})I_{4}(t_{2})\rangle}{T^2}.
\label{C34}
\end{equation}
This is a measurable quantity, where $I_{j}(t)$ are current operators at ports $j=3,4$. These are given by the sum of electron (e) and hole (h) current operators along the corresponding terminals
\begin{eqnarray}
I_{3}(t)&=&I^{(\mathrm{e})}_{L1}(t)+I^{(\mathrm{h})}_{L2}(t), \\
I_{4}(t)&=&I^{(\mathrm{e})}_{R1}(t)+I^{(\mathrm{h})}_{R2}(t).
\end{eqnarray}
More precisely, the hole currents are defined as
\begin{eqnarray}
I^{(\mathrm{h})}_{R2}(t)&=&-I^{(\mathrm{e})}_{R2}(t)+I^{(\mathrm{e})}_{0},\\
I^{(\mathrm{h})}_{L2}(t)&=&-I^{(\mathrm{e})}_{L2}(t)+I^{(\mathrm{e})}_{0},
\end{eqnarray}
where $I^{(\mathrm{e})}_{0}$ is the mean electronic current in either terminal L2 or R2 when BS-0 is \emph{closed} (see Fig.~\ref{fig-3}), namely, when no electron-hole pairs are emitted. Moreover, the electron current operators at terminals $n={\mathrm L1},{\mathrm L2},{\mathrm R1},{\mathrm R2}$ are defined as\cite{Buttiker-92}
\begin{equation}
I^{(\mathrm{e})}_n(t)=\frac{e}{h\nu}\sum_{E,\omega}e^{-i\omega t}b^\dag_n(E)b_n(E+\hbar\omega).
\end{equation} 
In Eq.\ (\ref{C34}), $T$ is the measurement time and $\nu$ is the
density of states of the leads (where we consider a discrete spectrum to
ensure a proper regularization of the current correlations). The
expectation value $\langle\cdots\rangle$ is taken over the probe state
$|\Psi\rangle$ given in Eq.~(\ref{input}). Notice that, since the redefined vacuum 
$|\bar{0}\rangle$ correspond to a noiseless stream of electrons emitted from
BS-2, $\langle\bar{0}|I_{3}(t_{1})I_{4}(t_{2})|\bar{0}\rangle=0$. In
addition, it can be shown that
$\langle\bar{\Psi}|I_{3}(t_{1})I_{4}(t_{2})|\bar{0}\rangle=\langle\bar{0}|I_{3}(t_{1})I_{4}(t_{2})|\bar{\Psi}\rangle=0$. As a consequence, we find
\begin{equation}
\langle\Psi|I_{3}(t_{1})I_{4}(t_{2})|\Psi\rangle=\langle\bar{\Psi}|I_{3}(t_{1})I_{4}(t_{2})|\bar{\Psi}\rangle,
\end{equation}
which brings us to focus our attention on the state
$|\bar{\Psi}\rangle$ of Eq.~(\ref{Psi_bar}). This is normalized by calculating
$\langle\bar{\Psi}|\bar{\Psi}\rangle$, which involves a double energy integral. Fermionic algebra reduces the expression to 
\begin{eqnarray}
\langle\bar{\Psi}|\bar{\Psi}\rangle&=&|t_{0}|^2\int_{0}^{eV} d\varepsilon \int_{0}^{eV} d\varepsilon^{\prime}
\delta_{\epsilon,\epsilon^\prime}\times\nonumber\\
\big(|t_{1}|^2|t_{2}|^2&+&|r_{1}|^2|r_{2}|^2+|r_{1}|^2|t_{2}|^2+|r_{2}|^2|t_{1}|^2\big).
\end{eqnarray} 
The terms within brackets sum to one by unitarity. Moreover, $\int_{0}^{eV}{\mathrm
d}\varepsilon=\mathcal{N}=eV\nu$ is the number of energy states in the
leads. This leaves
$\langle\bar{\Psi}|\bar{\Psi}\rangle=\mathcal{N}|t_{0}|^{2}$, which is
nothing but the tunnel current through BS-0 in units of $e/\nu h$.\cite{note-2} We
now introduce the normalized state
\begin{equation}
|{\bar{\Psi}'}\rangle=\frac{1}{\sqrt{\mathcal{N}}|t_{0}|}|\bar{\Psi}\rangle
\end{equation}
which can be written, up to a global phase factor, as 
\begin{eqnarray}
\label{Psi_bar_norm}
|{\bar{\Psi}'}\rangle&=&|t_{1}||r_{2}||\Phi_{{\mathrm L}{\mathrm L}}\rangle+|r_{1}||t_{2}||\Phi_{{\mathrm R}{\mathrm R}}\rangle \nonumber\\
&+&\sqrt{|t_{1}|^{2}|t_{2}|^{2}+|r_{1}|^{2}|r_{2}|^{2}}|\Phi_{{\mathrm L}{\mathrm R}}\rangle.
\end{eqnarray}
Here, $|\Phi_{ij}\rangle$ ($i,j={\mathrm L},{\mathrm R}$) are normalized states defined as
\begin{eqnarray}
|\Phi_{{\mathrm L}{\mathrm L}}\rangle&=&\frac{t_{0}}{\sqrt{\mathcal{N}}|t_{0}|}\int_{0}^{eV} d\varepsilon\frac{t_{1}r^{\star}_{2}e^{i(\gamma_{1}-\gamma_{2})}}{|t_{1}||r_{2}|}a^{\dagger}_{\mathrm L1}(\varepsilon)a_{\mathrm L2}(\varepsilon)|\bar{0}\rangle \nonumber\\
|\Phi_{{\mathrm R}{\mathrm R}}\rangle&=&\frac{t_{0}}{\sqrt{\mathcal{N}}|t_{0}|}\int_{0}^{eV} d\varepsilon\frac{r_{1}t^{\star}_{2}e^{i(\delta_{1}-\delta_{2})}}{|r_{1}||t_{2}|}a^{\dagger}_{\mathrm R1}(\varepsilon)a_{\mathrm R2}(\varepsilon)|\bar{0}\rangle \nonumber\\
|\Phi_{{\mathrm L}{\mathrm R}}\rangle&=&\frac{t_{0}}{\sqrt{\mathcal{N}}|t_{0}|}\int_{0}^{eV} d\varepsilon\frac{1}{\sqrt{|t_{1}|^{2}|t_{2}|^{2}+|r_{1}|^{2}|r_{2}|^{2}}}\nonumber\\
&&[t_{1}t^{\star}_{2}e^{i(\gamma_{1}-\delta_{2})}a^{\dagger}_{\mathrm L1}(\varepsilon)a_{\mathrm R2}(\varepsilon) \nonumber\\
&-&r_{1}r^{\star}_{2}e^{-i(\gamma_{2}-\delta_{1})}a_{\mathrm L2}(\varepsilon)a^{\dagger}_{\mathrm R1}(\varepsilon)]|\bar{0}\rangle.
\label{Phi-states}
\end{eqnarray}
%
\begin{figure}
\includegraphics[width=\columnwidth]{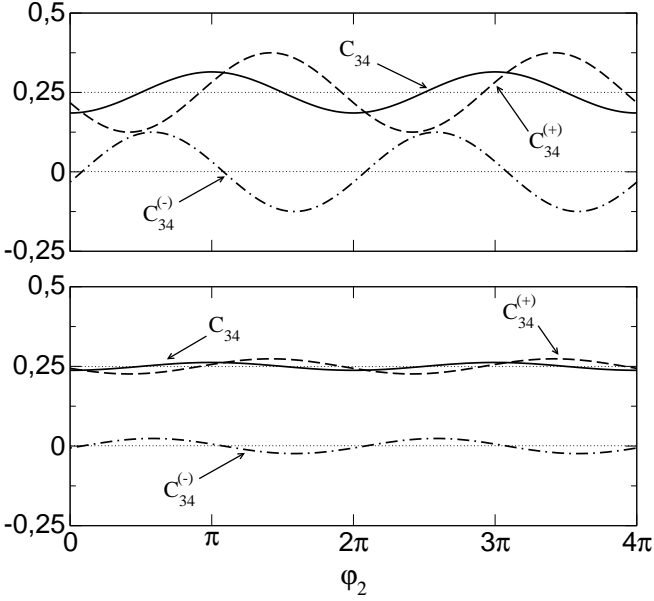}
\caption{Entanglement witness signals as a function of the controllable phase $\varphi_2$.  A $\mathcal{C}_{34} < 1/4$ (solid line) is indicative of  hyper-entanglement, while spatial-mode and occupancy entanglement are discriminated by $\mathcal{C}^{(+)}_{34} < 1/4$ (dashed line) and $\mathcal{C}^{(-)}_{34} \neq 0$ (dash-dotted line), respectively. Top panel: $50\%$ BS-1 and BS-2 ($t_1=r_1=t_2=r_2=1/\sqrt{2}$), where large-amplitude oscillations demonstrate highly entangled components. Bottom panel: $95\%$ transparent BS-1 and BS-2 ($t_1=t_2=\sqrt{19/20}$ and $r_1=r_2=1/\sqrt{20}$), where low-amplitude oscillations are indicative of weakly entangled components (similar results, not shown here, are obtained for low-transparency or highly asymmetric  settings of BS-1 and BS-2). Phases are set to $\gamma_1=\gamma_2=\delta_1=\delta_2$ and $\varphi_1=5\pi/12$ without any loss of generality (alternative settings would only introduce relative phase shifts in the curves).}
\label{fig-4}
\end{figure}
%
The $|\Phi_{{\mathrm L}{\mathrm R}}\rangle$ corresponds to a delocalized L-R component of the electron-hole excitation contributing to spatial-mode qubit entanglement (addressed by postselection in Ref.~\onlinecite{FC-09}) while $|\Phi_{{\mathrm L}{\mathrm L}}\rangle$ and $|\Phi_{{\mathrm R}{\mathrm R}}\rangle$ contribute to occupancy entanglement. 
By introducing $\theta,\phi\in[0,\pi/2]$ such that \cite{note-3}
\begin{eqnarray}
\sin\theta&=&\sqrt{|r_{1}|^{2}|t_{2}|^{2}+|r_{2}|^{2}|t_{1}|^{2}}, \nonumber\\
\sin\phi&=&\frac{|r_{1}||t_{2}|}{\sqrt{|r_{1}|^{2}|t_{2}|^{2}+|r_{2}|^{2}|t_{1}|^{2}}},
\end{eqnarray}
the state in Eq.~(\ref{Psi_bar_norm}) reduces to
\begin{equation}
|\bar{\Psi}'\rangle=\sin\theta(\cos\phi|\Phi_{{\mathrm L}{\mathrm L}}\rangle+\sin\phi|\Phi_{{\mathrm R}{\mathrm R}}\rangle)+\cos\theta|\Phi_{{\mathrm L}{\mathrm R}}\rangle.
\label{Psinorm}
\end{equation}
This has the exact form of the generic two-particle input state
considered in Ref.~\onlinecite{GFTF-07}. By following this
reference, we can write
\begin{equation}
\mathcal{C}_{34}=\big[1+w\cos^{2}\theta+v\sin^{2}\theta\sin(2\phi)\big]/4,
\label{C34a}
\end{equation}
where $w$ and $v$ real quantities satisfying $|w|,|v| \le 1$ such that $0 \le \mathcal{C}_{34} \le 1/2$. These depend on the controllable phases $\varphi_{\alpha}$ and the numerical coefficients appearing in states
$|\Phi_{ij}\rangle$ (\ref{Phi-states}) denoted as $\Phi^{(ij)}_{\alpha,\beta}$, where $\alpha,\beta=1,2$ refer to 
the electron/hole propagating channels. More precisely, in this case we arrive to
\begin{eqnarray}
w&=&\int_{0}^{eV} d\varepsilon\big[(\Phi^{({\mathrm L}{\mathrm R})}_{12})^{\ast}\Phi^{({\mathrm L}{\mathrm R})}_{21}e^{i(\varphi_{1}-\varphi_{2})}+\nonumber\\
&&(\Phi^{({\mathrm L}{\mathrm R})}_{21})^{\ast}\Phi^{({\mathrm L}{\mathrm R})}_{12}e^{i(\varphi_{2}-\varphi_{1})} \big],\\
v&=&-{\mathrm Re}\Big\{\int_{0}^{eV} d\varepsilon\big[(\Phi^{({\mathrm L}{\mathrm L})}_{12})^{\ast}\Phi^{({\mathrm R}{\mathrm R})}_{12}e^{i(\varphi_{1}+\varphi_{2})}\big]\Big\}.
\end{eqnarray}
From this, together with Eq. \ (\ref{Phi-states}), the current cross correlator $\mathcal{C}_{34}$ of Eq. (\ref{C34a}) reduces to
\begin{eqnarray}
\mathcal{C}_{34}&=&\frac{1}{4}(1-2\mathrm{Re}[r_{1}t^{\star}_{1}r^{\star}_{2}t_{2}e^{i(\delta_{2}-\gamma_{1}+\delta_{1}-\gamma_{2}+\varphi_{1}-\varphi_{2})}]\nonumber\\
&-&2\mathrm{Re}[r_{1}t^{\star}_{1}r_{2}t^{\star}_{2}e^{i(-\delta_{2}-\gamma_{1}+\delta_{1}+\gamma_{2}+\varphi_{1}+\varphi_{2})}]).
\label{C34b}
\end{eqnarray}
This correlator has been shown\cite{GFTF-07} to work as a hyper-entanglement witness: values of $\mathcal{C}_{34}$ smaller that 1/4 for some setting of the controllable phases $\varphi_1$ and  $\varphi_2$ indicate the non-separability of $|\bar{\Psi}'\rangle$ in the L-R bipartition, representing a direct evidence of entanglement in the probe state without revealing its specific (either mode or occupancy) form. Moreover, entanglement-specific witnesses can be defined by appropriate data processing as $\mathcal{C}^{(\pm)}_{34} = [\mathcal{C}_{34}(\{\varphi_\alpha\}) \pm \mathcal{C}_{34}(\{\varphi_\alpha+\pi/2\})]/2$, finding\cite{GFTF-07} 
\begin{eqnarray}
\label{C34plus-0}
\mathcal{C}^{(+)}_{34} &=& (1+w \cos^2 \theta)/4, \\
\mathcal{C}^{(-)}_{34} &=& v \sin^2\theta \sin(2 \phi)/4,
\label{C34minus-0}
\end{eqnarray}
such that $0 \le \mathcal{C}^{(+)}_{34} \le 1/2$ and $-1/4 \le \mathcal{C}^{(-)}_{34} \le 1/4$.
The particular expressions for the interferometer of Fig.~\ref{fig-3}, derived form Eq. (\ref{C34b}), reduce to
\begin{eqnarray}
\mathcal{C}^{(+)}_{34}&=&\frac{1}{4}-\frac{1}{2}\mathrm{Re}[r_{1}t^{\star}_{1}r^{\star}_{2}t_{2}e^{i(\delta_{2}-\gamma_{1}+\delta_{1}-\gamma_{2}+\varphi_{1}-\varphi_{2})}] \label{C34plus},\\
\mathcal{C}^{(-)}_{34}&=&-\frac{1}{2}\mathrm{Re}[r_{1}t^{\star}_{1}r_{2}t^{\star}_{2}e^{i(-\delta_{2}-\gamma_{1}+\delta_{1}+\gamma_{2}+\varphi_{1}+\varphi_{2})}].
\label{C34minus}
\end{eqnarray}
The presence of spatial-mode qubit entanglement in the probe state is
revealed whenever $\mathcal{C}^{(+)}_{34}<1/4$, corresponding to negative values of $w$. \cite{GFTF-06,GFTF-07} Occupancy
entanglement, instead, manifests as a $\mathcal{C}^{(-)}_{34}$
different from zero. From Eqs.~(\ref{C34plus}) and (\ref{C34minus}) we
notice that these conditions are satisfied for some values of the
controllable phases $\varphi_1$ and $\varphi_2$ provided the
transmission ($t_1,t_2$) and reflection ($r_1,r_2$) amplitudes of beam
splitters BS-1 and BS-2 are non-vanishing (i.e., for \emph{partly}
open quantum point contacts). The witness signals are optimized for $50\%$ BS-1 and BS-2. However, hyper-entanglement is a constraint impeding the saturation of the algebraic bounds allowed by Eqs. \ (\ref{C34plus-0}) and (\ref{C34minus-0}): that would require either $\cos^2\theta=1$ in Eq. \ (\ref{C34plus-0}) [corresponding to a purely LR component in Eq. (\ref{Psinorm})] or $\sin^2\theta=1$ and $\sin 2\phi=\pm1$ in Eq. \ (\ref{C34minus-0}) [no LR component and balanced LL and RR ones in Eq. \ (\ref{Psinorm})], something impossible to accomplish in our setup.

For illustration, in Fig.~\ref{fig-4} we plot $\mathcal{C}_{34}$
(solid line), $\mathcal{C}^{(+)}_{34}$ (dashed line) and
$\mathcal{C}^{(-)}_{34}$ (dash-dotted line) for two different BS-1 and
BS-2 transparencies as a function of the controllable phase
$\varphi_2$. For simplicity, we set
$\gamma_1=\gamma_2=\delta_1=\delta_2$ and $\varphi_1=5\pi/12$.  The top
panel corresponds to $50\%$ BS-1 and BS-2, showing large-amplitude
oscillations indicative of highly entangled spatial-mode and occupancy
components. Still, the amplitudes do not saturate due to hyper-entanglement as pointed out above. The bottom panel shows results for highly transmitting
($95\%$ transparency) BS-1 and BS-2. Low-amplitude oscillations are a
signal of partial entanglement, eventually undetectable depending on
the experimental accuracy. Notice that similar results are obtained
for either low transparencies or highly asymmetric settings of BS-1
and BS-2 since the oscillation amplitudes in Fig.~\ref{fig-4} are
fully determined by the products $r_1t_1^*r_2t_2^*$ and $r_1t_1^*r_2^*t_2$
according to Eqs. (\ref{C34b}), (\ref{C34plus}) and (\ref{C34minus}).

To conclude, we find that $\mathcal{C}^{(+)}_{34}$ sets a lower bound to the concurrence\cite{W98} $C$ of the two spatial-mode qubits. We first notice from Ref. \onlinecite{GFTF-07} that the entanglement of formation\cite{EoF} $E_{\rm f}$ of the state $|\Phi_{\rm LR}\rangle$ satisfies $E_{\rm f}(\Phi_{\rm LR}) \ge \mathcal{E}(1-2 \mathcal{C}_{34}(\Phi_{\rm LR}))$, with $\mathcal{E}(x)$ a monotonically increasing function for $1/2 \le x \le 1$ (null otherwise) \cite{note-4} and $\mathcal{C}_{34}(\Phi_{\rm LR})= (1+ w)/4$ the correlator of Eq. (\ref{C34a}) for $\cos \theta = \pm 1$ (i.e., when the state (\ref{Psinorm}) carries only a delocalized L-R component contributing exclusively to spatial-mode qubit entanglement). Moreover, the two-qubit entanglement of formation and the concurrence are related in the form $E_{\rm f} = \mathcal{E}((1+C)/2)$.\cite{W98} This already implies that $C \ge 1- 4 \mathcal{C}_{34}(\Phi_{\rm LR})$. Finally, by noticing that $\mathcal{C}_{34}(\Phi_{\rm LR}) \le \mathcal{C}^{(+)}_{34}$ for $w \le 0$ we find that 
\begin{equation}
C \ge 1- 4 \mathcal{C}^{(+)}_{34}. 
\end{equation}
The concurrence runs from $0$ for separable two-qubit states to $1$ for maximally entangled (Bell) states. Hence, a $0 \le \mathcal{C}^{(+)}_{34} < 1/4$ is indicative of some degree of spatial-mode qubit entanglement. In particular, a vanishing $\mathcal{C}^{(+)}_{34}$ would be a sign of maximally entangled states. In our setup, however, $\mathcal{C}^{(+)}_{34}$ can not vanish due to the constraints imposed by hyper-entanglement (see above). Moreover, a $1/4 \le \mathcal{C}^{(+)}_{34}$ sets a negative lower bound to $C$ meaning that no claims can be done about the entanglement under such particular circumstances.

\section{Closing remarks}

Our proposal integrates a reliable electronic entangler with a versatile analyzer, basis of an entanglement witness qualified to discriminate spatial-mode and occupancy entanglement with limited resources at reach.\cite{NOCHMU07} This includes the ability to quantify the entanglement by appropriate lower bounds. \cite{GFTF-07} The witness is particularly suitable since it is optimized to detect the precise family of states that the entangler is able to produce. This means that, in ideal condition and in contrast to most witnesses, a negative signal (i.e., vanishing oscillation amplitudes in Fig.~\ref{fig-4}) is indicative of  \emph{no entanglement}. Moreover, the witness remains sound even in the presence of noisy inputs in the form of mixed states, as demonstrated in Ref.~\onlinecite{GFTF-07}. 

\acknowledgments

We acknowledge support from projects Nos. P07-FQM-3037 (CEIC, Junta de Andaluc\'{\i}a), FIS2011-29400 and FIS2014-53385-P (MINECO, Spain) with FEDER funds.



\begin{thebibliography}{99}
\bibitem{note-1}
For the sake of direct comparison with the interferometer introduced in Ref.~\onlinecite{FC-09}, we keep the terminology in terms of the original left and right elements of the interferometer.
\bibitem{note-2}
Despite the state $|\bar{\Psi}\rangle$ is a two-particle (electron-hole) excitation, its norm $\langle\bar{\Psi}|\bar{\Psi}\rangle$ is proportional to a current (and not to a square current) because the electron-hole pair is created in a single tunneling event.
\bibitem{note-3}
These trigonometric functions are well defined thanks to the unitarity of the scattering matrices of BS-1 and BS-2.
\end{thebibliography}
\end{document}